\documentclass[epj]{svjour}
\usepackage{graphics}
\usepackage[all]{xy}
\usepackage{mathabx}

\newcommand{\bb}{\bibitem}
\newcommand{\bes}{\begin{subequations}}
\newcommand{\ees}{\end{subequations}}
\newcommand{\ben}{\begin{eqnarray}}
\newcommand{\een}{\end{eqnarray}}
\newcommand{\be}{\begin{equation}}
\newcommand{\ee}{\end{equation}}
\newcommand{\sech}{\textrm{sech}}

\newcommand{\arcsinh}{\textrm{arcsinh}}
\begin{document}
\title{Study of models of the sine-Gordon type in flat and curved spacetime}
\author{D. Bazeia\inst{1,2,3} \and L. Losano\inst{2,3} \and R. Menezes\inst{3} \and Rold\~ao da Rocha\inst{4}% etc
}                     % Do not remove
\offprints{}          % Insert a name or remove this line
\institute{Instituto de F\'\i sica Universidade de S\~ao Paulo, 05314-970 S\~ao Paulo SP, Brazil. \and Departamento de F\'\i sica Universidade Federal da Para\'{\i}ba, 58051-900 Jo\~ao Pessoa PB, Brazil. \and Departamento de F\'\i sica Universidade Federal de Campina Grande, 58109-970, Campina Grande PB, Brazil. \and Centro de Matem\'atica, Computa\c c\~ao e Cogni\c c\~ao, Universidade Federal do ABC 09210-170, Santo Andr\'e, SP, Brazil.}
\date{Received: date / Revised version: date}
\abstract{We study a new family of models of the sine-Gordon type, starting from the sine-Gordon model, including the double sine-Gordon, the triple one, and so on. The models appears as deformations of the starting model, with the deformation controlled by two parameters, one very small, used to control a linear expansion on it, and the other, which specifies the particular model in the family of models. We investigate the presence of topological defects,  showing how the solutions can be constructed explicitly from the topological defects of the sine-Gordon model itself. In particular, we delve into the double sine-Gordon model in a braneworld scenario with a single extra dimension of infinite extent, showing that a stable gravity scenario is admissible. Also, we briefly show that the deformation procedure can be used iteratively, leading to a diversity of possibilities to construct families of models of the sine-Gordon type. 
\PACS{
      {PACS-key}{11.10. Lm, 11.27. +d}  
      % \and
      %{PACS-key}{discribing text of that key}
     } % end of PACS codes
} %end of abstract
\maketitle
\section{Introduction}

Topological defects appear in nature in a diversity of contexts. They are of current interest in several branches of nonlinear science, in particular in high energy and condensed matter physics, where they can be used to describe phase transitions in the early universe, map interfaces separating distinct regions and contribute to pattern formation in nature \cite{r1,VS,MS,DW}. An interesting recent example regarding the use of topological defects  concerns the study of magnetic domain wall in a nanowire, intended for the development of magnetic memory \cite{AV}. Other investigations of current interest, concerning topological defects in high energy physics, can be found for instance in Refs.~\cite{Dunne,spain}.

In this work we study the presence of topological defects in relativistic models described by a real scalar field $\phi$ in $(1,1)$ spacetime dimensions. We focus attention on the deformation procedure, a method introduced in \cite{BLM1}, and further used to describe the presence of topological defects in models described by scalar fields in a diversity of scenarios; see, e. g., Refs. \cite{ABLM,ASD,Epl1}. In particular, in \cite{Epl1} one investigated a new family of sine-Gordon models, engendering interesting features.  

The results obtained in Refs.~\cite{BLM1,ABLM,ASD,Epl1} inspired us to study the new family of models, which we introduce in the current work. Here we start with the sine-Gordon model, and subsequently  a deformation function controlled by two parameters shall be used. One parameter is  very small, inducing a small deviation from the identity and the other allowing to specify the member in the family of models. A direct advantage of this procedure is that the implementation is straightforward, leading to potentials of direct interest to practical applications. Another advantage is that the small parameter prominently helps us to find the topological solutions and the corresponding energy densities associated to the deformed models explicitly, in the several distinct topological sectors of the models.

The family of  models introduced in this work can be used in a diversity of contexts, in particular, to study issues suggested in Ref.~\cite{sine}. Another route of current interest concerns the investigation of braneworld models \cite{Brane} with distinct solutions. The sine-Gordon model was used in \cite{G} to generate a stable braneworld model, and this could guide us to investigate the issue when the scalar field describes a particular member in the family of models  introduced in this work. See also \cite{K,D}, for other recent studies on this model in warped spacetime. For pedagogical motivation, the work is organized  as follows: in Sec.~II we introduce the necessary background for the present study. In the sequel, we implement the deformation procedure and construct the family of models in Sec.~III. In Sec.~IV we study the double sine-Gordon model in curved space time, in a braneworld model with a single extra dimension of infinite extent. There we focus on the profile of the warp factor, the energy density and stability of the model.

We close the work with comments and conclusions in Sec.~V. There we briefly show how to use the composed deformation, in which one uses the same deformation to deform the deformed model once again. This iterative procedure changes the periodic behavior of the sine-Gordon model into a new, quasiperiodic behavior for the deformed model.

%%%%%%%%%%%%%%%%%%%%%%%%%%%%%%%%%%%%%%%%%%%%
\section{Generalities}

We start considering models described by the Lagrange density
\be\label{lagrangian}
{\cal L}=\frac12 \partial_\mu \chi \partial^\mu \chi - V(\chi),
\ee
where $V(\chi)$ is the potential which specifies the model under consideration, described by the real scalar field $\chi=\chi(x,t)$. Furthermore, $x^\mu=(x^0=t, x^1=x)$. For simplicity, we are using natural units, and we have rescaled the field, the space and time coordinates as well, to make them dimensionless.

Some basic facts about topological solutions are briefly reviewed in what follows, focusing our attention on static fields. The equation of motion for $\chi=\chi(x)$ is given by
\be
\frac{d^2\chi}{dx^2}=\frac{dV}{d\chi}.
\ee
Topological solutions usually appear when the potential is non-negative, which can be written in the following form, using $W=W(\chi)$, 
\be\label{neruzhy}
V(\chi)=\frac12 W_\chi^2,
\ee
where $W_\chi$ stands for $dW/d\chi$. In this case one gets
\be
\frac{d^2 \chi}{dx^2}=W_\chi W_{\chi \chi}.
\ee
The point here is that this equation can be solved by solutions of the first order equation
\be\label{firstorder}
\frac{d\chi}{dx}=W_\chi.
\ee
Since the potential does not realize the sign of $W$, Eq. (\ref{firstorder}) can also be seen with $W$ changed to $-W$. The approach then maps the second order equation of motion into two first order equations, which in general holds for topological solution. 

A topological solution, which solves the first order equation on the topological section ($jk$), has energy minimized to the value $E^{jk}_{BPS}=|W(\bar\chi_j)-W(\bar\chi_k)|$, where $\bar \chi_j$ and $\bar \chi_k$ are two adjacent minima in the set $ \{ \bar\chi_1, \bar\chi_2, \ldots, \bar\chi_n \}$ of minima related to the model. This is the Bogomol'nyi bound, and the solution of the first order equation is named BPS state \cite{BPS}, which is linearly stable. Since $W(\chi)$ is a function of $\chi$, it can be used to define the topological behavior, which can be seen from the topological current 
\be\label{topologic}
j_T^\mu = \varepsilon^{\mu\nu} \partial_\nu W(\chi).
\ee
This definition for the topological current is different from the standard form, which uses $j^\mu = \varepsilon^{\mu\nu} \partial_\nu \chi$; see, e. g., \cite{r1}. Notwithstanding, one can use (\ref{topologic}) to show that the topological charge of the solution equals its energy, apart from a sign factor, and is more appropriate in general \cite{Dbazeia}.

\subsection{Deformation Procedure}

We focus attention on the deformation procedure proposed in \cite{BLM1}. An important advantage of using this approach is that we can in a straightforward way get new models and construct their topological defects analytically. Here we briefly review the procedure, starting with the standard model (\ref{lagrangian}). 
The model is supposed to support topological defects and as well as the corresponding solutions are known analytically.  In this paper, we investigate small modifications to the sine-Gordon model whose potential is given by 
\be\label{cosss}
V(\chi)=\frac12 \cos^2 (\chi).
\ee
This potential is obtained from the function $W(\chi)=\sin(\chi)$ and have an infinite number of minima $\bar \chi = \pm n \pi/2$, where $n=1,3,5,\ldots$ The topological solutions that connect those minima are
\be\label{chi_x}
\chi_s(x)=\pm (\theta(x)+k\pi),
\ee
where $k=0,\pm1,\pm2,\ldots$ and $ \theta(x)=\arcsin(\tanh(x))$ is the Gudermannian function. The energy is $E=2$, since we are using dimensionless units. 

According to the deformation procedure, we can consider another model, described by
\be
{\cal L}=\frac12 \partial_\mu \phi \partial^\mu \phi - U(\phi).
\ee
Here $U(\phi)$ is the new potential, which is written in terms of the starting potential $V(\chi)$ as
\be
U(\phi)=\frac{V(\chi \to f(\phi))}{f^{\prime2}(\phi)},
\ee
where $f(\phi)$ is the deformation function. In this case, if $\chi(x)$ is a static solution of the starting model, then we get $\phi(x)$ as a solution of the new, deformed model, which is given by
\be
\phi(x)=f^{-1}[\chi(x)].
\ee
This is the general procedure, and now we specialize to the case where the deformation function describes small deviation from the identity. 

Let us choose $f(\phi)=\phi+\epsilon g(\phi)$, where $\epsilon$ is very small parameter to be used to allow power expansion on it. However, we should do it with care, since the function $g(\phi)$ cannot increase too much to break the approximation up to the first order power on $\epsilon$,  which shall be implemented hereon.

Although the deformation procedure holds  for a general potential, let us concentrate on models controlled by $W$, described by Eq.(\ref{neruzhy}). 
In this case we get that
\be\label{calWp}
U(\phi)=\frac12 {\cal W}_\phi^2,
\ee
where the new ${\cal W}={\cal W}(\phi)=W(\phi)+\epsilon W_\epsilon(\phi)$ is such that
\be
{\cal W}_\phi = W_\phi - \epsilon \left( W_\phi g_\phi - W_{\phi\phi} g\right).
\ee
Thus
%%%%%%%%%%%%%%%%%%%%%%%
\be
W_\epsilon(\phi)=-\int^\phi d\phi\, \left( W_\phi g_\phi - W_{\phi\phi} g\right)
\ee

%%%%%%%%%%%%%%%%%%%%%%%

This is general result, and we can obtain the minima of the new potential from the equation
\be
W_\phi=\epsilon \left( W_\phi g_\phi - W_{\phi\phi} g\right).
\ee 
The minima can be used to find the corresponding energy in each topological sector, in the form $E_{ij}=|{\cal W}_i-{\cal W}_j|$, where ${\cal W}_i$ and ${\cal W}_j$ stand for ${\cal W}(\phi_i)$ and ${\cal W}(\phi_j)$, respectively, with $\phi_i$ and $\phi_j$ representing two adjacent minima of the new potential. Also, from the expression which defines the deformation function we can write, up to the first order in $\epsilon$, the solution of the new model as $\phi=\chi-\epsilon g(\chi)$, where $\chi$ represents a solution of the starting, sine-Gordon model. 

%%%%%%%%%%%%%%%%%%%%%%%%%
\section{New Family of Models}

To introduce the new family of models, let us consider the starting model as the sine-Gordon given by Eq. (\ref{cosss}). As one knows, this model is an important model, and the above procedure will deform it, leading to other models which are close to the original model. Moreover, since we are starting from a periodic potential, one can naturally choose $g(\phi)$ periodic, making it limited in a such a way that the first order approximation in $\epsilon$ remains valid in the full real line. In this way, our main goal is to introduce new family of models, as the double and the triple sine-Gordon models. 
%%%%%%%%%%%%%%%%
\begin{figure}\sidecaption\resizebox{0.9\hsize}{!}
{\includegraphics*{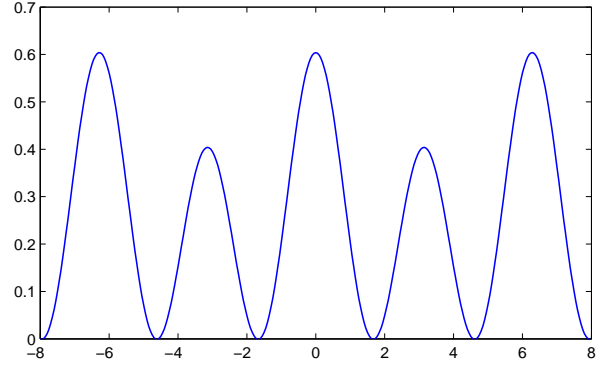}}
\caption{Potential of the double sine-Gordon model, depicted for $s=1$ and $\epsilon=-0.1$}.
\end{figure} 
%%%%%%%%%%%%%%%%%%
Here we take the deformation function in the form
\be\label{funcdeform}
f_s(\phi)=\phi+\epsilon \sin\left(\frac{\phi}{s}\right),
\ee
parametrized by the two parameters, $s$ and $\epsilon$, real, with $\epsilon$ very small. Using the sine-Gordon model as the starting model, this function leads to the model described by the potential (\ref{calWp}), where ${\cal{W}}$ is now such that
\begin{equation}\label{wsg}
{\cal W}_\phi \!= \!\cos(\phi) - \!\epsilon\! \left[\sin(\phi)\sin\!\left(\frac{\phi}{s}\right)\!+\frac{1}{s}\! \cos(\phi)\cos\!\left(\frac{\phi}{s}\right)\!\right].
\end{equation}
The additional term that arises shifts the minima of the potential to
\be\label{minima}
\phi^{min}_\pm = \pm \left(\frac{n\pi}{2}-\epsilon \sin \left(\frac{n\pi}{2s}\right)\right), \,\,\,\,\, n=1,3,5,\ldots
\ee
It also changes the maxima
\be\label{maxima}
\phi^{max}_\pm = \pm \left(m\pi + \frac{\epsilon(s^2-1)}{s^2}\sin\left(\frac{m\pi}{s}\right)\right), \,\,\,\,\, m=0,1,2,\ldots
\ee
and their respective heights
\be
h_{m+1}=\frac12 -\frac{\epsilon}{s} \cos\left(\frac{m_s \pi}{s}\right), \,\,\,\,\, m_s=0,1,2,\ldots, s.
\ee
The modified potential is realized to have periodicity $2\pi s$. 

%%%%%%%%%%%%%%%%%%%%
\begin{figure}\sidecaption\resizebox{0.9\hsize}{!}
{\includegraphics*{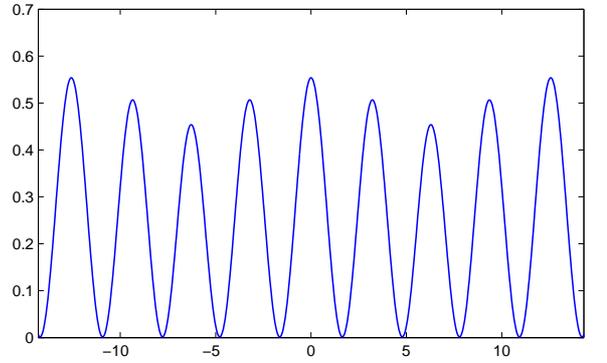}} 
\caption{Potential of triple sine-Gordon model, depicted for $s=2$ and $\epsilon=-0.1$}.
\end{figure} 
%%%%%%%%%%%%%%%%%%%%%%%%%%%%%

For $s$ integer, we see that the number of distinct topological sectors, which are labeled by $m$, is given by $s+1$. Then, for $s=1$ we get to the double sine-Gordon, for $s=2$, the triple sine-Gordon model, and so on. The family of models which is generated in the present work is different from the family introduced in Ref.~\cite{Epl1}, since there we started with another model, and used a different deformation function. To see how the models behave in present case, in Figs. 1 and 2 the potentials are depicted for $s=1$ and $s=2$, respectively. 

For $s\neq 1$, the new function ${\cal W}$  can be written as 
\ben
{\cal W}= \sin(\phi) \!&-&\! \frac{\epsilon}{2}\Bigg[\frac{(s+1)}{(s-1)}\sin \left(\frac{(s-1)}{s}\phi\right) \nonumber \\
&&-\frac{(s-1)}{(s+1)} \sin \left(\frac{(s+1)}{s}\phi\right) \Bigg],
\een
and for $s=1$
\be
{\cal W}=\sin(\phi) - \epsilon \phi.
\ee

We can use ${\cal W}$ and the minima given by Eq. (\ref{minima}), to calculate the energy of the topological defect in each one of the topological sectors of the respective model. The general expression for $\cal W$ in each minimum is 
\be\label{Wsuper}
{\cal W}(\phi^{min}_\pm)=\pm (-1)^{\frac{n-1}{2}} \left(1 - \frac{2s}{s^2-1} \cos\left( \frac{n\pi}{2s} \right)\epsilon\right),
\ee
and for $s=1$
\be
{\cal W}(\phi^{min}_\pm)=\pm\left( (-1)^{\frac{n-1}{2}} -\epsilon \frac{n\pi}{2}\right).
\ee
%%%%%%%%%%%%%%%%%%%%%%%%%
\begin{figure}\sidecaption\resizebox{0.9\hsize}{!}
{\includegraphics*{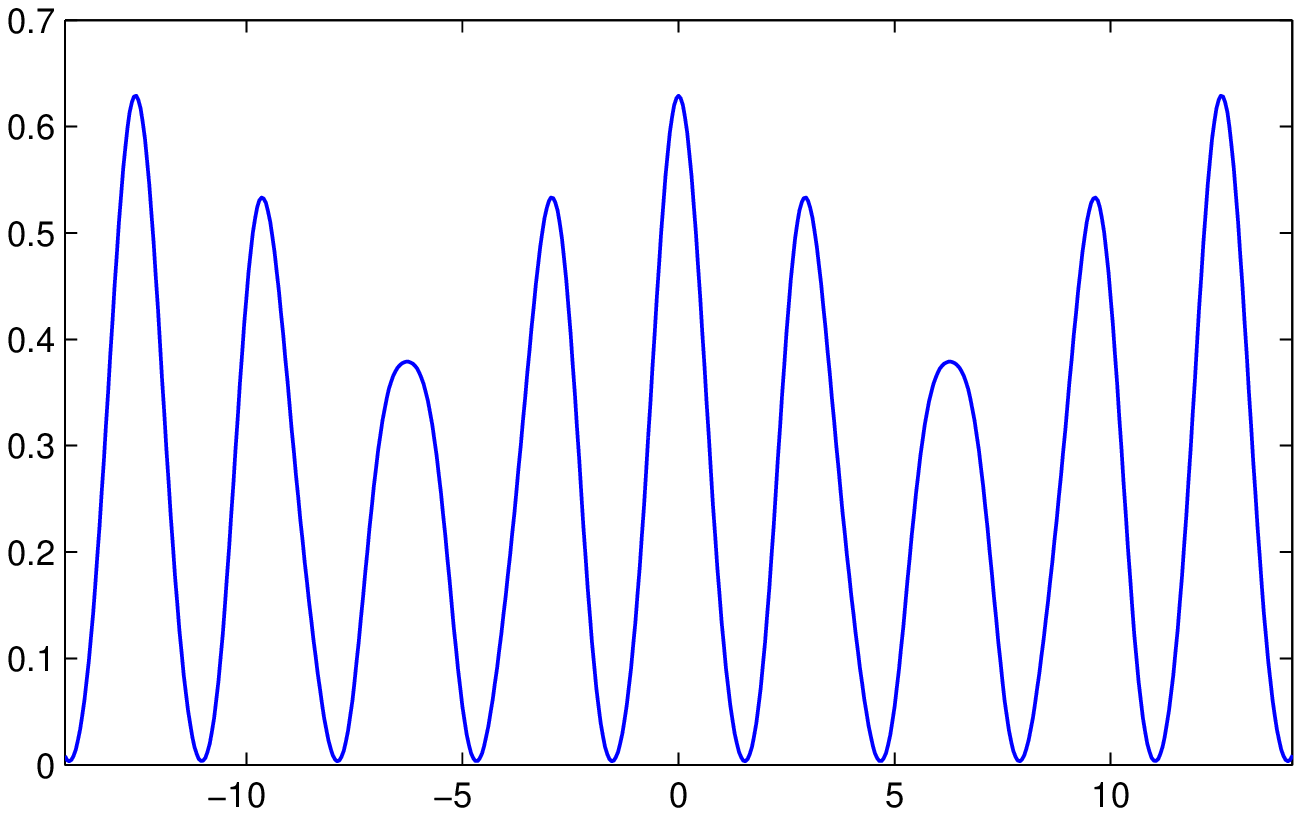}}
\caption{Potential associated to the generalized sine-Gordon model, for $s=2/5$ and $\epsilon=-0.05$.}
\label{Fig3}
\end{figure} 
%%%%%%%%%%%%%%%%%%%%%%%%

%%%%%%%%%%%%%%%%%%%%%%%%
\begin{figure}\sidecaption\resizebox{0.9\hsize}{!}
{\includegraphics*{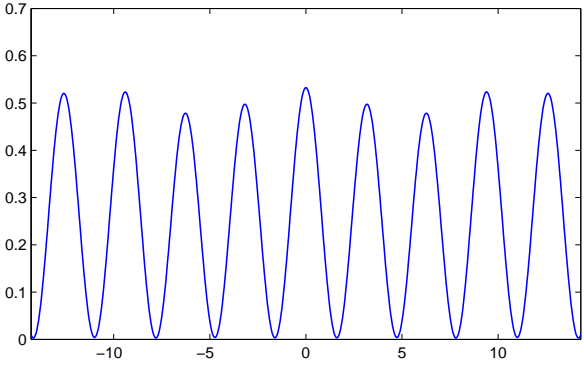} }
\caption{Potential associated to the generalized sine-Gordon model, for $s=\sqrt{3}$ and $\epsilon=-0.05$.}
\label{Fig4}
\end{figure} 
%%%%%%%%%%%%%%%%%%%%%%%%%%%%%%%

Moreover, the topological solutions associated to the first order equation (\ref{firstorder}) can be obtained from the inverse of the deformation function. Eq. (\ref{funcdeform}) is hence used to get
\be
\phi(x)=\phi_0(x)-\epsilon \sin\left(\frac{1}{s} \phi_0(x)\right),
\ee
where $\phi_0(x)$ represents the solutions of the sine-Gordon model, given by Eq. (\ref{chi_x}); recall that $\phi_0(x)=\chi_s(x)$. With this, it is not hard to calculate the corresponding energy density, which is given by
\be
\rho(x)=\rho_0(x) \left[1-\frac{2\epsilon}{s} \cos\left(\frac{\phi_0(x)}{s}\right) \right], 
\ee
with $\rho_0(x)=\sech^2(x)$ being the energy density of the corresponding solutions. 

Let us now consider the case of the triple sine-Gordon model, which is obtained with $s=2$. In this case, there are three kinds of topological sectors. The sectors of the first kind are described by the following solutions
\be
\phi_1(x)=\theta(x)+4l\pi - \epsilon\, {\rm sgn}(x) \sqrt{\frac{1}{2} \left(1-\sech(x)\right)},
\ee
where ${\rm sgn}(x)$ is the sign function and $l$ is an integer that connect the minima
\be
-\frac{\pi}{2}+4l\pi + \epsilon \, \frac{\sqrt{2}}{2}, \,\,\, \quad{\rm and}\,\,\,\,\quad \,\,\;\frac{\pi}{2}+4l\pi - \epsilon \, \frac{\sqrt{2}}{2}.
\ee
The sectors of the second kind are described by the solutions
\be
\phi_2(x)=\theta(x) +(4l+1) \pi - \epsilon \, \sqrt{\frac{1}{2} \left(1+\sech(x)\right)},
\ee
which connect the minima 
\be
\frac{\pi}{2}+4l\pi - \epsilon \, \frac{\sqrt{2}}{2},\quad\; \,\,{\rm and}\,\, \,\,\,\;\quad\frac{3\pi}{2}+4l\pi - \epsilon \, \frac{\sqrt{2}}{2}.
\ee
The sectors of the third kind are described by the solutions
\be
\phi_3(x)=\theta(x) + 2l\pi + \epsilon \, {\rm sgn}(x) \sqrt{\frac{1}{2} \left(1-\sech(x)\right)},
\ee
which connect the minima 

\be
\frac{3\pi}{2}+4l\pi - \epsilon \, \frac{\sqrt{2}}{2}, \,\,\, {\rm and}\,\, \frac{5\pi}{2}+4l\pi + \epsilon \, \frac{\sqrt{2}}{2}.
\ee

The energy densities of the solutions corresponding to the three distinct topological sectors are given by

\be
\rho_1(x)=\left(1-\epsilon\, \sqrt{\frac{1}{2} \left(1+\sech(x)\right)}\right) \,\sech^2(x),
\ee
\be
\rho_2(x)=\left(1+\epsilon\, {\rm sgn}(x) \sqrt{\frac{1}{2} \left(1-\sech(x)\right)}\right) \,\sech^2(x),
\ee
\be
\rho_3(x)=\left(1+\epsilon\, \sqrt{\frac{1}{2} \left(1+\sech(x)\right)}\right) \,\sech^2(x).
\ee

We use Eq. (\ref{Wsuper}) to obtain the corresponding energies
\be
E_1=2-\frac{4\sqrt{2}}{3} \,\epsilon, \,\,\,\, E_2=2, \,\,\,\, E_3=2+\frac{4\sqrt{2}}{3}\, \epsilon.
\ee

There are many other values for $s$ which introduce new features to the deformed models. The specific sequence $s=2/3, \,1/2, \, 2/5, \,1/3,\,\ldots,$ leads to interesting potentials, and to other kinds of sine-Gordon models. To illustrate this fact, in Fig. 3 we depict the potential for $s=2/5$. The topological solutions and their amplitudes, widths, energy densities and energies can all be calculated straightforwardly, using the approach shown above. One can also take $s$ irrational, leading to quasiperiodic potentials. An illustration of this is given in Fig. 4, for $s=\sqrt{3}$. As before, the topological solutions, amplitudes, widths, energy densities and energies can also be calculated straightforwardly. 

%%%%%%%%%%%%%%%%%%%
\section{Double sine-Gordon model in warped spacetime}

Let us now focus on the double sine-Gordon model in a warped spacetime. 
From the deformation procedure proposed heretofore, we aim to study and unravel the stability of  the associated braneworld scenario. It is known that  given a scalar field model by the potential associated to it, it is possible to compute the warp factor \cite{Brane}. Therefore, the braneworld stability and the graviton zero mode can be analyzed. As the braneworld stability is known for the sine-Gordon model \cite{G}, we aim here to study whether deformations of such model, induced by the small parameter, preserves both the stability and the masslessness of the graviton. We investigate the double sine-Gordon model, which is the simplest in the family of new models introduced in the current work.

Here we follow previous studies \cite{G,K,D} and, in order to simplify the calculations, we represent the double sine-Gordon model with
\be
W(\phi)={2a^2}\sin(\phi/a); \,\,\,\,\, g(\phi)=a \cos(\phi/a),
\ee
where $a$ is a real parameter, to be fixed below.
This choice leads us to
\ben\label{dsG3}
{\cal W}(\phi) &=& W(\phi) + \epsilon W_\epsilon (\phi)\nonumber\\
&=&{2a^2}\sin(\phi/a)-2\epsilon {a\phi}.
\een
We consider this model in a braneworld scenario with a single extra dimension of infinite extent, governed by the action
\be
S=\int d^5x\sqrt{|g|}{\left(-\frac14 R+ \frac12 \partial_\mu\phi\partial^\mu \phi-V(\phi)\right)},
\ee
where $4\pi G=1$. The scalar fields and the warp factor are assumed to depend merely on the extra dimensional coordinate $y$. The potential has the form
\be\label{potbrane}
V(\phi)=\frac18\; {\cal W}^2_\phi- \frac13 {\cal W}^2,
\ee
and the metric is controlled by
\be
ds^2=e^{2A(y)}\eta_{\mu\nu}dx^\mu dx^\nu-dy^2,
\ee
which represents the standard $AdS_5$ metric. As one knows, when the scalar field depends solely upon the extra dimension, the equations of motion 
\ben
\phi^{\prime\prime}+4A^\prime\phi^\prime&=&
\frac{dV}{d\phi};\qquad\quad
A^{\prime\prime}=-\frac23\,\phi^{\prime2}
\\
A^{\prime2}&=&\frac16\phi^{\prime2}-\frac13 V(\phi),
\een
reduce to the two first-order equations
\be\label{WWp}
\frac{d\phi}{dy}=\frac12 {\cal W}_\phi;\;\;\;\;\; \frac{dA}{dy}=-\frac13 {\cal W}(\phi).
\ee

%%%%%%%%%%%%%%%%%%%%%%%%%%%%%%%%%%

The potential given by Eq. (\ref{potbrane}) can be written as
\be
V(\phi)=\frac12 W_\phi^2 -\frac13 W^2 + \epsilon \left[ W_\phi W_{\epsilon\phi} - \frac23 W W_{\epsilon}\right].
\ee
Using (\ref{dsG3}), we get 
\ben
V(\phi)&=&-\frac{5a^2}{12} +\frac{11a^2}{12} \cos(2\phi/a) \nonumber\\
&&-{4a^3\epsilon } \left[\frac23 \sin(\phi/a)-\frac{\phi}{a} \cos(\phi/a) \right].
\een
In this expression, the term depending on $\epsilon$ may increase too much for large values of $\phi$, and this may invalidate our calculation up to first-order in $\epsilon$. To avoid problem with this, let us suppose that the scalar field is also small, varying around $\phi=0$. Therefore, we focus attention to the central sector of the model, where the defect solution is given by 
\ben
\phi(y)&=& \phi_0(y) -\epsilon g(\phi_0)\\
&=&a\arcsin(\tanh(y)) -{\epsilon a} \tanh(y)
\een

Manipulating the above Eqs. (\ref{WWp}), we can express the warp function in terms of the scalar field
\be
A(\phi)=-\frac23\int^\phi d\phi \,\frac{{\cal W}}{{\cal W}_\phi} ,
\ee
or yet
\be
A(\phi)\!=\!-\frac23\int^\phi \!\!d\phi \frac{W}{W_\phi} -\frac{2\epsilon}{3}\int^\phi\!d\phi\left(\frac{W_\epsilon}{W_\phi}-\frac{WW_{\epsilon\phi}}{W_\phi^2}\right) .
\ee
Substituting $\phi(y)=\phi_0(y)-\epsilon g(\phi_0(y))$, we get
\ben
A(\phi)   &&=-\frac23 \int^{\phi_0} d\phi \,\frac{W}{W_\phi} \\&&+\frac{2\epsilon}3 \left[\frac{W(\phi_0)}{W_\phi(\phi_0)} g(\phi_{0}) - \!\int^{\phi_0} \!\!d\phi\,\left(\frac{W_\epsilon}{W_\phi}\!-\!\frac{WW_{\epsilon\phi}}{W_\phi^2}\right) \!\right] .\nonumber
\een
We use (\ref{dsG3}) to write
\ben
A(\phi_0)&=&-\frac{2a^2}{3}\ln (\cosh(\phi_0/a)) \nonumber \\&&-\frac{2 a^2\epsilon}{3}\!\left[ \cos(\phi_0/a)\!-\!\frac1{a^2}\!\!\int^{\phi_0}\!\!\!\!\! d\phi\,\phi\,\sec(\phi/a)\right]\!\! .\;\;\;
\een
In the central sector of the model, the scalar field is small and the defect solution has the form 
\be
\phi_0(y)=a\arcsin(\tanh(y)).
\ee
We substitute it and we find the warp factor
\ben\label{Ay1}
A(y)=\frac{2a^2}{3} \ln (\sech(y)) - \frac{2a^2\epsilon}{3} \left[\sech(y) - L(y)\right],
\een
where 
\be
L(y)=\int^{\phi_0(y)}\!\!\!\!\! d\phi \,\phi \sec(\phi).
\ee
Here we have set to zero an integration constant that corresponds to an irrelevant shift in the warp factor $A$. Therefore we choose $L(0)=1$. For $y \to \pm \infty$ we have
$A(y)\sim - (2/3) b^{-2}|y|$; thus, asymptotically the warp factor does not depend on $\epsilon$.  Up to first-order in $\epsilon$, the warp factor is 
\be
e^{2A(y)}=\sech{^{\frac{4a^2}{3}}}(y)\left[ 1 + \frac{2a^2\epsilon}{3} (\sech(y) - L(y))
\right].
 \ee

In order to examine stability regarding  the braneworld scenario, the metric is perturbed as
\be
ds^2=e^{2A(y)}(\eta_{\mu\nu}+h_{\mu\nu})dx^\mu dx^\nu-dy^2,
\ee 
where $h_{\mu\nu}=h_{\mu\nu}(x,y)$ represent small perturbations.
We choose $h_{\mu\nu}$ in the form of transverse and traceless contributions denoted by ${\bar h}_{\mu\nu}$  \cite{Brane}; it follows that 
\be\label{h}
{\bar h}_{\mu\nu}^{\prime\prime}+4\,A^{\prime}
\,{\bar h}_{\mu\nu}^{\prime}=e^{-2A}\,\Box\,{\bar h}_{\mu\nu},
\ee
where  $\Box$ denotes the Lorentzian Laplace-Beltrami operator. This choice makes gravity to decouple from the matter field.
We now use the conformal coordinate $dz=e^{-A(y)}dy$, and also ${\bar h}_{\mu\nu}(x,z)=e^{ik\cdot x}e^{-\frac{3}{2}A(z)}H_{\mu\nu}(z)$;
in this case, Eq. (\ref{h}) becomes the Schr\"odinger-like equation
\be\label{se}
-\frac{d^2H_{\mu\nu}}{dz^2}+V(z)\,H_{\mu\nu}=k^2\,H_{\mu\nu},
\ee
with the associated potential 
\be\label{potquant}
V(z)=\frac32\,A^{\prime\prime}(z)+\frac94\,A^{\prime2}(z).
\ee
As the Hamiltonian in Eq.~(\ref{se}) can be expressed as $
H=\left(-\frac{d}{dz}-\frac32A^{\prime}\right)
\left(\frac{d}{dz}-\frac32A^{\prime}\right)$, it implies  that $k$ is real, and $k^2\geq0$. We then conclude that no negative energy bound state is present, so the brane is stable.

The conformal coordinate $z$ is written as
\ben
z&=&\int dy \, {\cosh^{\frac{2a^2}{3}} (y)} \\&&+ \frac{2a^2\epsilon}{3} \int dy \, \left[(\cosh(y))^{\frac{2a^2}{3}-1} - (\cosh(y))^{\frac{2a^2}{3}} L(y)\right].\nonumber
\een
We choose $a^2=3/2$ to obtain
\be\label{zbyy}
z=\sinh(y) +\epsilon \left(y - {\sinh(y)} L(y) + \frac{3}{2}M(y) \right)
\ee
with
\ben
M(y)\!=\!\frac16\int^{\phi_0} \! d\phi \,\phi\, \tan(\sqrt{{2}/{3}}\phi)\sec(\sqrt{2/3}\phi), 
\een
and $M(0)=0$. We invert (\ref{zbyy}) to get
\be\label{yyy}
y(z)=\arcsinh(z) + \epsilon y_\epsilon (z),
\ee
where 
\ben\label{yepsilon}
y_\epsilon(z)&=&-\frac{1}{\sqrt{z^2+1}} \bigg[\arcsinh(z)\\ &&-z L\left({\arcsinh(z)}\right)+\frac32 M\left(\arcsinh(z)\right)\bigg] .\nonumber
\een
We now substitute $y=y(z)$ in (\ref{Ay1}) in order to obtain
\ben
A(z)=-\frac12 \ln(1+z^2)+ \epsilon A_\epsilon (z),
\een
where $A_\epsilon$ can be found from (\ref{yyy}) and (\ref{Ay1}). 

\begin{figure}\sidecaption\resizebox{0.9\hsize}{!}
{\includegraphics*{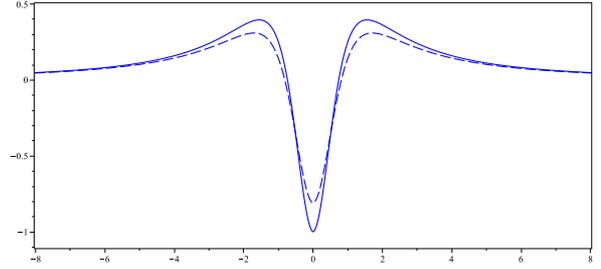}}
\caption{The volcano potential $V(z)$, associated to the Schr\"odinger-like equation (\ref{se}), depicted with solid and dashed curves, for $\epsilon=0$ and $-0.1$, respectively.}
\label{pq}
\end{figure}

The associated potential (\ref{potquant}) is given by
\be
V(z)=  \frac{3}{4} \frac{(5z^2-2)}{(1+z^2)^2} + \epsilon\, V_\epsilon (z),
\ee
with
\be
V_\epsilon=\frac{3}{2}A^{\prime\prime}_\epsilon+  \frac94  \frac{ z}{1+z^2} A^{\prime}_\epsilon.
\ee
We depict in Fig.~\ref{pq} the potential $V(z)$ for $\epsilon=0$ and $-0.1$. It is of the volcano type, and the zero mode is the only bound state,
thus ensuring stability and masslessness of the graviton.

%%%%%%%%%%%%%%%%%%%%%%%%
%%%%%%%%%%%%%%%%%%%%%%%%%%%%%%%

%%%%%%%%%%%%%%%%%%%
\section{Comments and conclusions}

The quasiperiodic behavior of the potential can be better seen if one changes the deformation function appropriately. Actually, we note that the same deformation function can be subsequently used to deform the deformed model afresh. Since $\epsilon$ is small parameter, if one uses $g_1(\phi)$ for the first deformation, and $g_2(\phi)$ for the second one, one ends up with the deformation 
\be\label{secondit}
f(\phi)=\phi+\epsilon (g_1(s_1)+g_2(s_2)).
\ee
Note that the procedure can be repeated iteratively, evoking interesting deformations. Here, however, we only discuss the second iteration, which is explicitly given by Eq. (\ref{secondit}). If the sine-Gordon potential is deformed with the function (\ref{secondit}), we get to a diversity of interesting potentials.  A typical example of this depicted in Fig. 6, for $s_1=1$ and $s_2=2$. Another interesting case is given by $s_1=\sqrt{2}$ and $s_2=\sqrt{3}$, which is depicted in Fig. 7. This last case enhances the quasiperiodic behavior of the potential in a significant way, as commented above. As before, all solutions can be obtained, and the corresponding amplitudes, widths, energy densities and energies straightforwardly. 

%%%%%%%%%%%%%%%%%%%%%%%%%%%%%%%%%
\begin{figure}\sidecaption\resizebox{0.9\hsize}{!}
{\includegraphics*{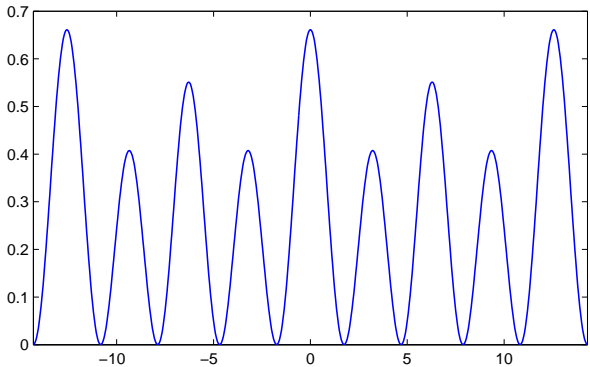}} 
\caption{Potential regarding the generalized sine-Gordon model, for $s_1=1$, $s_2=2$ and $\epsilon=-0.1$}.
\end{figure}
%%%%%%%%%%%%%%%%%%%%%%%%

%%%%%%%%%%%%%%%%%%%%%%%%%% 
\begin{figure}
\sidecaption\resizebox{0.9\hsize}{!}
{\includegraphics*{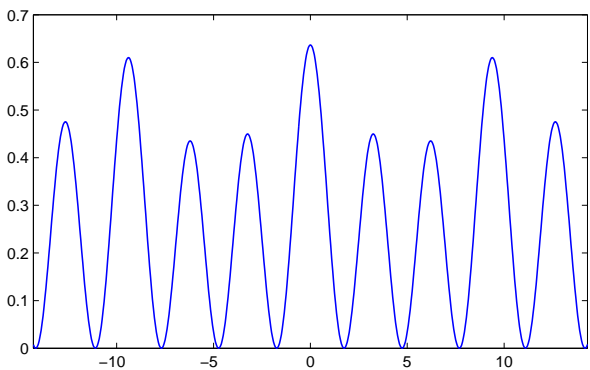}} 
\caption{Potential regarding the generalized sine-Gordon model, for $s_1=\sqrt{2}$, $s_2=\sqrt{3}$ and $\epsilon=-0.1$}.
\end{figure} 
%%%%%%%%%%%%%%%%%%%%%%%%%

The procedure implemented in Sec.~III engenders a diversity of possibilities, with all topological defects being explicitly constructed from the topological defects of the sine-Gordon model. The solutions, and their amplitudes, widths, energy densities and energies are straightforward to be obtained. The models can be studied in a simple way, and the suggested family of models can be used for practical applications in a diversity of scenarios, in particular in the braneworld context in five dimensions, with a single extra dimension of infinite extent, with the brane being stabilized by a real scalar field \cite{Brane,G,K,D}. A study on this was introduced in Sec.~IV, where the double sine-Gordon model was investigated in the standard $AdS_5$ spacetime.

Since the procedure involves a small deformation of a given model, it can be used in a more general sense, to study small deformations of other models, of interest to high energy physics and to other branches of nonlinear science. For instance, we can start from the double sine-Gordon model, instead of the sine-Gordon, to generate another family of models. Moreover,  the same methodology can be applied to more general models, including the case where the dynamics of the scalar field is modified, to deal with higher order power in the first derivative of the field, which is also of current interest to high energy physics. In particular, the presence of higher order power in the derivative of the field can be used to the study of twinlike theories \cite{t1}, which are distinct models having the same defect solutions, with the very same energy density. For models with generalized dynamics, we can also study the presence of compactons, which are defect solutions living in a closed, compact interval of the real line \cite{c1}. Some of these issues are presently under consideration, and will be reported elsewhere.

\section*{Acknowledgments}
The authors would like to thank C. A. G. Almeida, M. Pazetti and J. G. G. S. Ramos for comments and suggestions, and CAPES, CNPq and FAPESP for partial financial support. R. da Rocha thanks to CNPq grant 303027/2012-6.


\begin{thebibliography}{}
\bibitem{r1} R. Rajaraman, {\it Solitons and Instantons} (North-Holland, Amsterdan, 1982).
\bibitem{VS} A. Vilenkin and E. P. S. Shellard, {\it Cosmig Strings and Others Topological Defects} (Cambridge, UK, 1994).
\bibitem{MS} N. Manton and P. Sutcliffe, {\it Topological Solitions} (Cambridge UP, Cambridge, UP, Cambridge, UK, 2004).
\bibitem{DW} D. Walgraef, {\it Spatio-temporal Pattern Formation} (Springer-Verlag, New York, 1997).
\bibitem{AV} A. Vanhaverbeke, A. Bischof and R. Allenspach, Phys. Rev. Lett. {\bf 101}, 107202 (2008).
\bb{Dunne}G. Basar and G.V. Dunne, Phys. Rev. Lett. {\bf100}, 200404 (2008).
\bb{spain}A. Alonso-Izquierdo, M.A. Gonzalez Leon, and J. Mateos Guilarte, Phys. Rev. Lett. {\bf101}, 131602 (2008).
\bibitem{BLM1} D. Bazeia, L. Losano, and J. M. C. Malbouisson, Phys Rev. D {\bf 66}, 101701 (2002).
\bibitem{ABLM} C. A. Almeida, D. Bazeia, L. Losano, and J. M. C. Malbouisson, Phys. Rev. D {\bf 69}, 067702 (2004); V. I. Afonso, D. Bazeia, M. A. Gonzalez Leon, L. Losano, and J. Mateos Guilarte, Nucl. Phys. B {\bf 810}, 427 (2009); D. Bazeia, Ashok Das, L. Losano, and M. J. Santos, Applied Math. Lett. {\bf 23}, 681 (2010).
\bibitem{ASD}A. de Souza Dutra, Physica D {\bf 238}, 798 (2009); W. T. Cruz, M. O. Tahim, and C. A. S. Almeida, Europhys. Lett. {\bf 88}, 41001 (2009); A. E. R. Chumbes and M. B. Hott, Phys. Rev. D {\bf 81}, 045008 (2010); R.R. Landim, G. Alencar, M.O. Tahim, R.N. Costa Filho, JHEP {\bf1108}, 071 (2011);
R. Landim, G. Alencar, M.O. Tahim, M.A.M. Gomes, R.N. Costa Filho, EPL {\bf97}, 20003 (2012); A. E. Bernardini and R. da Rocha,  Advances in High Energy Phys. {\bf 2013}, 304980  (2013).
\bibitem{Epl1}  D. Bazeia, L. Losano, R. Menezes, and M. M. Sousa, EPL {\bf 87}, 21001 (2009).
\bibitem{sine} G. Mussardo, V. Riva, and G. Sotkov, Nucl. Phys. B {\bf 87}, 548 (2005); G. Mussardo, Nucl. Phys. B {\bf 779}, 101 (2007); A. Kundu, Phys. Rev. Lett. {\bf 99}, 154101 (2007); L. Benfatto, C. Castellani, and T. Giamarchi, Phys. Rev. Lett. {\bf 99}, 207002 (2007); L. A. Ferreira, B. Piette, and W. J. Zakrzewski, Phys Rev. E {\bf 77}, 036613 (2007); M. Cadoni, Y. -X. Liu, L. -D. Zhang, L. -D. Zhang, and Y. -S. Duan, Phys. Rev. D {\bf 78}, 0650025 (2008); D. Bazeia, L. Losano, J. M. C. Malbouisson, and R. Menezes, Physica D {\bf 237}, 937 (2008); J. H. Al-Alawi and W. J. Zakrzewski, J. Phys. A {\bf 41}, 315206 (2008); L. A. Ferreira, W. J. Zakrzewski, JHEP {\bf1105}, 130 (2011). 
\bibitem{Brane} L. Randall and R. Sundrum, Phys. Rev. Lett. {\bf 83}, 4690 (1999); W. D. Goldberger and M. B. Wise, Phys. Rev. Lett. {\bf 83}, 4922 (1999); O. DeWolfe, D. Z. Freedman, S. S. Gubser, and A. Karch, Phys. Rev. D {\bf 62}, 046008 (2000).
\bb{G}M. Gremm, Phys. Lett. B {\bf478}, 434 (2000).
\bb{K}R. Koley and S. Kar, Class. Quantum Grav. {\bf22}, 753 (2005).
\bb{D}Y. Brihaye and T. Delsate, Phys. Rev. D {\bf86}, 024029 (2012).
\bibitem{BPS}  E. B. Bogomol'nyi, Sov. J. Nucl. Phys. {\bf 24}, 449 (1976); M. K. Prasad and C. M. Sommerfield, Phys. Rev. Lett. {\bf 35}, (1975) 760.
\bibitem{Dbazeia} D. Bazeia, Phys. Rev. D {\bf 60}, 067705 (1999).
\bibitem{t1}M. Andrews, M. Lewandowski, M. Trodden, and D. Wesley, Phys. Rev. D {\bf82}, 105006 (2010); D. Bazeia, J.D. Dantas, A.R. Gomes, L. Losano, and R. Menezes, Phys. Rev. D {\bf84}, 045010 (2011); C. Adam and J.M. Queiruga, Phys. Rev. D {\bf84}, 105028 (2011); D. Bazeia and R. Menezes, Phys. Rev. D {\bf84}, 125018 (2011);
C. Adam and J.M. Queiruga, Phys. Rev. D {\bf85}, 025019 (2012); D. Bazeia and J. D. Dantas, Phys. Rev. D {\bf85}, 067303 (2012).
\bibitem{c1}D. Bazeia, E. da Hora, R. Menezes, H.P. de Oliveira and C. dos Santos, Phys. Rev. D {\bf81}, 125016 (2010); D. Bazeia, A. S. Lob\~ao Jr, and R. Menezes, Phys. Rev. D {\bf86}, 125021 (2012).\end{thebibliography}
\end{document}